\documentclass{technojournal} 
  
\usepackage[utf8]{inputenc} 
\usepackage{graphicx}
\usepackage{tabu,hhline,multirow,tabularx}
\usepackage{subfigure,epsfig} 
\usepackage{algorithm} 
\usepackage{algorithmic}
\def\@captype{figure} 
 
\usepackage{newtxtext} 
\usepackage{pstool}

\title[Efficient LBM on GPUs for dense moving objects using immersed boundary condition] 
{Efficient LBM on GPUs for dense moving objects using immersed boundary condition} 
  
\correspinfo[Joel Beny]{joel.beny@gmail.com} 
  
\author[acc]{Joel Beny}{} 
\author[acc2]{Jonas Latt}{} 
  
\affiliation[acc]{Université de Genève, Geneva, Switzerland, Joel.Beny@etu.unige.ch} 
\affiliation[acc2]{Universit\'e de Gen\`eve, Geneva, Switzerland, Jonas.Latt@unige.ch} 
  

\newcommand{\ba}{\begin{align}} 
\newcommand{\ea}{\end{align}} 
  
\begin{document} 
\maketitle 
\begin{abstract} 
    There exists an increasing interest for using immersed boundary methods (IBMs) \citep{peskin} to model moving objects in computational fluid dynamics. Indeed, this approach is particularly efficient, because the fluid mesh does not require to be body-fitted or to adjust dynamically to the motion of the body. Frequently, IBMs are implemented in combination with the lattice Boltzmann methods (LBM) \citep{lbm}. They fit elegantly into the framework of this method, and yield impressive parallel performances. It has also become quite common to accelerate LBM simulations with the use of Graphics Processing Units (GPUs) \citep{tolke}, as the underlying algorithm adjusts naturally to the architecture of such platforms. It is not uncommon that speedups of an order of magnitude, or more, at equal financial cost or energy consumption are observed, as compared to classical CPUs. IBM algorithms are however more difficult to adapt to GPUs, because their complex memory access pattern conflicts with a GPU's strategy of broadcasting data to a large number of GPU cores in single memory accesses. In the existing literature, GPU implementations of LBM-IBM codes are therefore restricted to situations in which the immersed surfaces are very small compared to the total number of fluid cells \citep{lara}, as is often the case in exterior flow simulations around an obstacle. This assumption is however not valid in many other cases of interest, as for example the simulation of deformable red blood cells (RBCs), for which many authors have adopted the LBM and IBM as their method of choice. Indeed, RBCs fill the blood volume densely, and the immersed surfaces contribute substantially to the overall computational cost. 

We propose a new method for the implementation of a LBM-IBM on GPUs in the CUDA language \citep{cuda}, which allows to handle a substantially larger immersed surfaces with acceptable performance than previous implementations. For test purposes, we consider the case of one or multiple rotating propellers in a fluid. The method is applied to a direct-forcing flavor of the IBM as proposed by Ota, Suzuki, and Inamuro \citep{ota}. In this method, the surface is represented by a certain number of points with Lagrangian coordinates, which interact with the fluid within a given kernel through a force term. The algorithm is iterative, to achieve a consistent result given the effect of neighboring, overlapping kernels. Each iteration is split into two parts: (1) a fluid $\rightarrow$ surface data transmission, during which the flow velocity is interpolated onto the immersed surface, and (2) a surface $\rightarrow$ fluid data transmission, during which the surface acts onto the fluid through a force term to enforce the boundary condition. 
\end{abstract} 
  
\graphicspath{{./}}

\section{Introduction} 
  
The objective of this work is to propose a GPU implementation of a 3D fluid simulation with immersed objects, using the immersed boundary methods (IBMs), and to observe a significant speed up compared to a CPU implementation, even in situations in which the immersed objects are densely packed.
  
To solve the incompressible Navier-Stokes equations for the fluid, we use the lattice Boltzmann (LBM) algorithm \citep{lbm}, within the Bhatnagar-Gross-Krook (BGK) framework. Multiple good GPU implementations already exist for this model, and our work is based on the implementation of Adrien Python, which is part of the Palabos framework. 
  
The LBM simulations are carried out on a regular, static, Euclidean mesh. This kind of simulation is well adapted to GPU implementations, and massive accelerations are achieved through many-core parallelization. A natural way to implement LBM on GPU is to assign a GPU thread to each lattice cell. With this strategy, our tests show a speed up of roughly a factor 20 as compared to a solid CPU implementation (the Palabos code). In this comparison, a standard high-end CPU and GPU are used, and all cores of the CPU are exploited through MPI-parallelism. Building on this, the challenge of the present work consisted in the integration of a IBM model into the software framework while maintaining a substantial gain of performance of the GPU as compared to the CPU.
  
The immersed object is represented by a set of Lagrangian points, which are superimposed to the Euclidean fluid mesh. The interaction between these points and the fluid is implemented in terms of a direct-forcing IBM, as described by Ota, Suzuki and Inamuro \citep{ota}. This algorithm is difficult to parallelize, because it involves non-local operations, such as the computation of averaged quantities over the extent of the interaction kernel of one point. Another difficulty stems from the mismatch of memory representations between the fluid's Eulerian lattice and the object's Lagrangian points which makes coalesced memory accesses difficult \citep{memory}. 
  
Our approach is based on the work of Valero-Lara, Pinelli, and Prieto-Matias \citep{lara}, in which the LBM fluid iterations and the implementation of the IBM force are carried out in separate CUDA kernels, by adopting different types of memory traversal in either case. A certain number of optimisations are proposed to achieve acceptable cases. As expected, this strategy cannot achieve ideal levels of parallelism, because parts of the computation remain sequential, and the bandwidth of memory accesses is suboptimal. Nevertheless, the implementation can achieve a substantial speedup of approximately a factor 19 compared to a modern, multi-core CPU, and is capable of efficiently handling situations with a large number of densely packed immersed objects.
  
The article first presents LBM and IBM methods used, and the details the algorithm of the CUDA implementation. The final part presents test cases and numerical results.  
  
\section{Lattice Boltzmann method} 
  
The computational domain is represented by a regular, homogeneous lattice. The degrees of freedom of the model consist of the \emph{populations}, a group of 19 variables $f_0, f_1, \cdots,  f_{18}$ on each lattice cell. At the beginning of each iteration of the fluid model, the macroscopic variables density ($\rho$) and velocity ($U$) are computed from the populations on each lattice cell as follows:

\begin{equation}\label{eq:u}    
\begin{gathered}    
\rho = \sum\limits_{i} f_i \quad  
U = \frac{1}{\rho}\sum\limits_{i} e_i f_i.
\end{gathered}    
\end{equation}   
  
In these equation, we have 19 structural vector $e_0 , e_1 , \cdots, e_{18}$ that reflect the connection between a cell and one of its neighbors, such as for example the vector $[1, -1, 0]$. This choice of vectors leads to the so-called D3Q19 model, which is summarized in~\citep{lbm}.
  
The value of the populations $f_i,\, i=0\cdots 18$ are updated at each iteration in two phases, the streaming and the collision phase. In the collision phase, the populations are updated locally, without communication with any neighbor. We use the so-called BGK collision model (for the initial of its authors Bhatnagar, Gross, and Krook)~\citep{bgk}:
\begin{equation}\label{eq:col}    
f_i(t) \leftarrow f_i(t) + \frac{1}{\tau}(Eq(i, t) - f_i(t)).
\end{equation}    
The constant $\tau$ is a characteristic relaxation time, which relates to the fluid viscosity, and depends on the time and space resolution. The operator $Eq()$, for equilibrium, depends on all populations on the local cell and is computed as follows:
     
\begin{equation}\label{eq:eq}    
Eq(i,t) = t_i \big[ 1 + 3 e_i \cdot U(t) + \frac{9}{2}(e_i \cdot U(t))^2 - \frac{3}{2} U(t) \cdot U(t) \big].
\end{equation}  
   
In this equation, the 19 values $t_i, i=0\cdots 18$ are constants, which are listed in~\citep{lbm}, which can be understood as weighting factors that compensate for the fact that some of the lattice vectors $e_i$  are longer than others.  
\vskip 10pt  
In the second phase, the streaming, the populations $f_i$ are propagated to their neighbouring cells, in direction of the corresponding vector $e_i$:  
\begin{equation}\label{eq:stream}  
f_i([X,Y,Z] + e_i, t+1) =  f_i([X,Y,Z],t)  
\end{equation} 
\vskip 10pt  
This concludes the description of the fundamental fluid algorithm. However, to further account for the presence of immersed objects through the IBM, a forcing term is added to the method, which modifies the collision phase. The computation of the value of the force resulting from the IBM is described in detail in the next section. For the time being, we describe it as a general function

$$\mathit{immersedBoundary}(\mathit{object}, U) \rightarrow G$$

that depends on all Lagrangian coordinates of the points describing the surface of the immersed object $\mathit{object}$, and on the velocity in all cells of the lattice $U)$, and produces a force $G$ to be exerted on all cells of the lattice. To apply the obtained force to the fluid, it is multiplied by the relaxation time $\tau$, and added to the velocity used to computed the equilibrium function.
\vskip 10pt 
In summary, the following pseudo code describes the full lattice Boltzmann algorithm:
\begin{enumerate}   
\item Computation of $U$ and $\rho$ according to~\eqref{eq:u}.   
\item Computation of $G$, $G = immersedBoundary(object, U)$, as described in the next section.
\item Inclusion of the force into the velocity term for the computation of the equilibrium: $U^G(X,Y,Z) = U(X,Y,Z) + \tau G[X,Y,Z)$ on each lattice cell.   
\item Computation of the 19 equilibrium values $Eq(i, \rho(X,Y,Z), U^G(X,Y,Z))$ on each lattice cell according to~\eqref{eq:eq}. 
\item Execution of the collision phase according to~\eqref{eq:col}.   
\item Execution of the streaming phase according to~\eqref{eq:stream}.
\end{enumerate}   
  
\section{Immersed boundary method} 
\begin{figure}[H]   
\centering   
\includegraphics[width=100 mm]{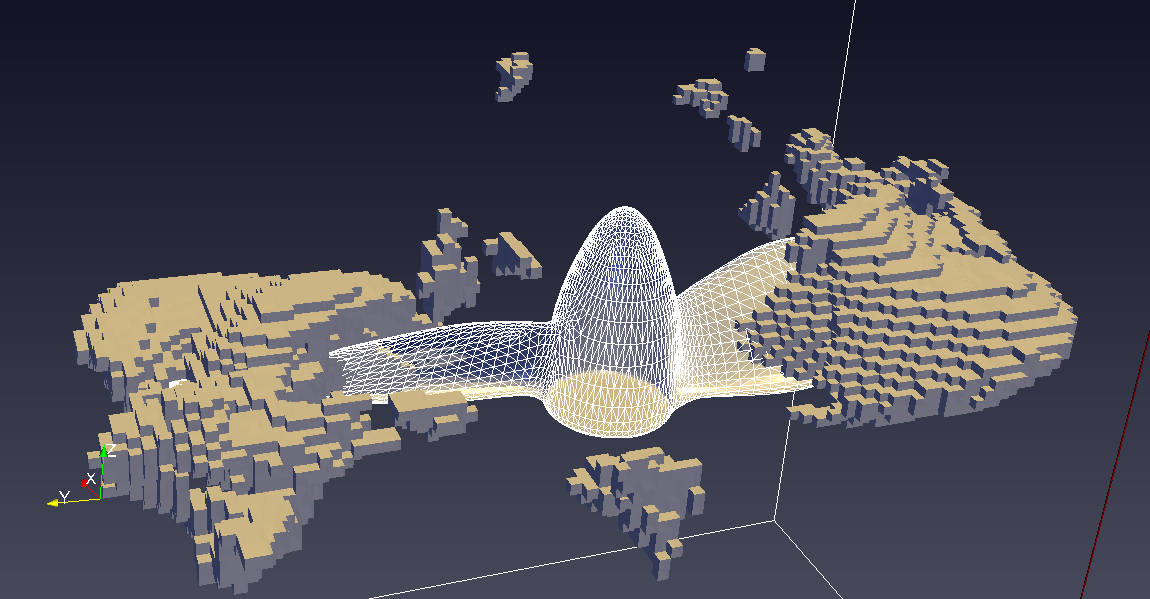}   
\caption{Interaction between regular fluid cells and an object described by a set of Lagrangian points.}   
\end{figure} 
The goal of the algorithm is to compute a force $G$ for each lattice cell which, once applied to the fluid, enforces a no-slip condition along the object surface. In case of a non-moving object, this amounts to enforcing a zero fluid velocity along the surface, and otherwise, a velocity equal to the local surface velocity of the object.
  
We will make a clear distinction between the velocity defined at the Eulerian positions of the lattice cells, described by a capital-letter $U(X,Y,Z)$, and the lowercase-letter velocity $u(k)$ defined at the Lagrangian positions on the surface of the obstacle. The first step of the immersed-boundary algorithm consists of the computation of the fluid velocity $u(k)$ at the Lagrangian positions $k$ on the surface of the object. It is computed as the weighted sum of the velocities $U$ of lattice cells in the neighborhood of the Lagrangian point:
$$ u_{iter0}(k) = \sum\limits_{X Y Z \in lattice} U_{iter0}(X,Y,Z)w(|x(k) - X| )w(|(y(k) - Y| )w(|z(k) - Z| )$$ 
The neighbourhood of the Lagrangian position $x(k)$ is defined by the weighting function $w(r)$. The result of this function, as applied to a given fluid cell, is proportional to the distance between the Lagrangian point and the fluid cell. It has as cut off for $r>2$, which means that the weighted sum is in practice computed only for a kernel of $5^3$ cells around the Lagrangian point.
\vskip 10pt 
We use the weighting function $w$ as defined by Peskin~\citep{peskin}: 
\[   
  w(r) =   
  \begin{cases}   
  \frac{1}{8}(3 - 2|r| + \sqrt{1 + 4|r|-4r^2})      & \text{if $r\leq 1$} \\   
  \frac{1}{8}(5 - 2|r| - \sqrt{1 - 7 + 12|r|-4r^2}) & \text{if $1\leq r\leq 2$} \\   
  0                                                 & \text{otherwise.}   
  \end{cases}   
\]  
  
As a next step, the force $g(k)$ exerted by the fluid on the object at the point $x(k)$ is computed as follows:
$$g_{iter0}(k)=(u_k - u_{iter0}(k)),$$    
where $u_k$ is the local velocity of the surface of the obstacle at position $x(k)$.
This provides a first estimate for $\Delta u(k)$, the velocity correction to be exerted on the fluid by the IBM:
$$\Delta u(k) = 0 + (u_k - u_{iter0}(k))s(k),$$     
where $s(k)$ is the area of the surface part represented by the Lagrangian point.

Finally, the fluid velocity is updated in the Eulerian mesh cells, to obtain a corrected field $U_{iter1}$ at iteration 1 as follows:
$$U_{iter1}(X,Y,Z) = U_{iter0}(X,Y,Z) + \sum\limits_{k \in object \: points} g_{iter0}(k)w(|(x(k) - X| )w(|y(k) - Y| )w(|z(k) - Z| ).$$    

At this point, the full procedure starts over, with the purpose to obtain a converged velocity field through a procedure of fixed-point iterations. As in the first iteration, a corrective force $g_{iter1}(k)$ is computed, leading to an updated value of the velocity correction $\Delta u(k)$:    
\begin{align*}    
& u_{iter1}(k) = \sum\limits_{X Y Z \in lattice} U_{iter1}(X,Y,Z)w(|x(k) - X| )w(|(y(k) - Y| )w(|z(k) - Z| ) \\    
\\    
& g_{iter1}(k)=(u_k - u_{iter1}(k)) \\    
&\Delta u(k) \leftarrow \Delta u(k) + (u_k - u_{iter1}(k))s(k)\\    
\\    
& U_{iter2}(X,Y,Z) = U_{iter1}(X,Y,Z) + \sum\limits_{k \in object \: points} g_{iter1}(k)w(|(x(k) - X| )w(|y(k) - Y| )w(|z(k) - Z| ) \\    
\end{align*}    
This procedure should in principle be repeated until convergence is reached, or in other terms, until the computed force correction $g(k)_{iter(n)}$ is negligibly small. In practice, we follow the recommendation by Ota, Suzuki, and Inamuro~\citep{ota} and apply the iterations $5$ times.

Finally, the correction $\Delta u(k)$ on the Lagrangian points is used to compute the force $G(X, Y, Z)$ to be applied to each lattice cell:    
$$G(X,Y,Z) = \sum\limits_{k \in object \: points}\Delta u(k)w(|(x(k) - X| )w(|y(k) - Y| )w(|z(k) - Z| ).$$
  
\section{CUDA implementation}  
CUDA is an application programming interface, working with the C, C++, and FORTRAN language, which allows us to run general-purpose code on NVidia GPUs. The CUDA API is articulated around function called \emph{kernels}, which are called from the CPU but executed on the GPU in a multi-threaded manner. Threads are grouped by blocks, which can have a three-dimensional shape. In this case, a kernel can refer to the current thread through 3 coordinates: \textit{threadIdx.x, threadIdx.y, threadIdx.z}. The blocks can themselves be arranged in a three-dimensional grid, using the three coordinates \textit{blockIdx.x, blokIdx.y blockIdx.z}. All in all, the current thread is referred to by 6 coordinates in a kernel.
  
Only GPU memory is accessible from within a thread. There exist 3 types of GPU memory. The first, global memory, is accessible by all threads, but is relatively slow, the second, shared memory, is shared among all threads of the same block and is significantly faster. The fastest type of GPU memory, are the registers, accessible only by the current thread and used for local variables.
  
Global memory accesses are most efficient if they are coalesced \citep{cuda}, meaning that neighbouring threads access neighbouring addresses. As an example, the following C-code instruction in a CUDA kernel
\begin{verbatim}
myGlobalArray[threadIdx.x] += 1
\end{verbatim}
represents a coalesced access, while the following does not
\begin{verbatim}
myGlobalArray[2*threadIdx.x] += 1
\end{verbatim}
  
The shapes of blocks and grids can be set up at the moment of a kernel call. The order of kernel calls are sequential by default, and the only way to synchronize code across different threads blocks is to proceed with subsequent kernel calls.
  
\subsection{Kernel calls} 
We divided each iteration of the LBM-IBM algorithm into different kernels for two reasons: firstly, to synchronize some part of the computation, and secondly, because we used different parallelization strategies and thread configurations for different parts of the algorithm.
  
In total, four kernels were used, which are
\begin{description}
    \item[The kernel \textit{lbm\_u\_start}] computes the fluid velocity $U$ from the populations $f_i$.
    \item The kernels \textbf{\textit{ib\_force1}} and \textbf{\textit{ib\_force2}} compute the force $G$ from the coordinates of the immersed object points and the fluid velocity $U$. This step has been split in two parts to allow synchronization of the threads after computation of the force $g(k)$ by \textit{ib\_force1} and the update of $U(X, Y, Z)$ and $G(X,Y,Z)$ in \textit{ib\_force2}.  
	\item[The kernel \textit{lbm\_computation}] computes the lattice Boltzmann collision and streaming phases depending on the populations of the previous step, and on $G$ and $U$. 
\end{description}
  
The kernels \textit{lbm\_u\_start} and \textit{lbm\_computation} are very closely related to the kernels described in Adrien Python's work~\citep{pyth} and are therefore not explained any further.
  
\subsection{The kernel \textit{ib\_force1}}
The purpose of this kernel is to compute the vectors $g(k)$ and $\Delta u(k)$ at each point $x(k)$ of the immersed object. Each given thread is responsible for the computation of one vector $g(k)$ and one vector $\Delta u(k)$ at a single Lagrangian point. A loop over neighboring lattice cells is carried out to access their velocities Eulerian velocities $U$. In this case, the corresponding memory accesses are often not coalesced. Indeed, fluid cells are arranged in memory according to a regular matrix ordering, row by row. As a consequence, they cannot be accessed consequently as a loop over the Lagrangian object positions and their Eulerian neighborhood is carried out, as it can be seen on Figure 2a. As it will be shown in the benchmark section, the impact of these non-coalesced accesses on the overall performance remains acceptable, as they correspond to relatively small memory chunks (the neighborhood of the solid surfaces), as compared to the overall size of the fluid domain. Furthermore, the performance impact of this kernel could be improved by reordering the Lagrangian positions on the object surface in the order of appearance of their nearest neighbor in the row-by-row matrix data structure, leading to an improved occurrence of coalesced accesses.
   
This kernel is also responsible for moving the points in the case of a non-static object, an operation that is fully performed within the GPU's memory to improve the performance. In this work, we focus entirely on rigid-body motion, as the one of a rotating propeller, and express therefore the motion as a linear transformation $M(t)$ applied to all points of the object. This transformation is applied at the first fixed-point iteration only. The updated positions are then stored and reloaded at the subsequent iterations, as they are needed to compute the point velocities $u(k)$.   
   
The vectors $g(k)$ and $\Delta u(k)$ are finally stored back in global memory, to be reused in subsequent kernel calls. These memory access have a negligible impact on performances because they are coalesced and are applied to a comparatively small amount of data.

The pseudo-code of this kernel has the following shape:
  
\begin{algorithm}[H] 
\caption{The kernel ib\_force1: computation of $g(k)$ and $\Delta u(k)$ on all object surface points $k$} 
\begin{algorithmic} 
\STATE ib\_force1(object, $U$, $M(t)$, fixed point iteration)\\ 
\STATE 
\STATE $k \gets blockDim.x \times blockIdx.x + threadIdx.x$\\ 
\IF{First fixed point iteration} 
    \STATE $x,y,z \gets $point $k$ rest position\\ 
    \STATE $\left( \begin{array}{c} 
    x(t)\\ 
    y(t)\\ 
    z(t) 
    \end{array} \right) 
    \gets M(t) \times  
    \left( \begin{array}{c} 
    x\\ 
    y\\ 
    z\\ 
    1 
    \end{array} \right)$\\ 
  
    \STATE $u_k \gets [x(t), y(t), z(t)] - [x(t-1), y(t-1), z(t-1)]$\\ 
  
\ELSE 
  
    \STATE $x(t),y(t),z(t) \gets $point $k$ transformed position\\ 
  
\ENDIF 
  
    \STATE $u(k) \gets [0,0,0]$\\ 
  
\FOR{each lattice cell $X, Y, Z$ in point $k$ neighbourhood} 
  
    \STATE $weight \gets w(|x(t)-X| ) \times w(|y(t)-Y|) \times w(|z(t)-Z|)$\\ 
    \STATE $u(k) \gets u(k) + U(X,Y,Z) \times weight$ 
\ENDFOR 
  
\STATE $g(k) \gets (u_k - u(k))$\\ 
  
\IF{First fixed point iteration} 
    \STATE $\Delta u(k) \gets 0$\\ 
\ELSE 
    \STATE $\Delta u(k) \gets \Delta u(k) + (u_k - u(k)) \times s(k)$\\ 
\ENDIF 
\end{algorithmic} 
\end{algorithm} 
  
\subsection{The kernel ib\_force2}
This kernel computes the corrected velocity $U$ and the force $G$ for each cell of the lattice. 
  
The parallelization strategy is fundamentally different to the one adopted in \textit{ib\_force1}, as a Lagrangian position $x(k)$ is assigned to a full CUDA block, and all threads of the block load the same values $\Delta u(k)$ and $g(k)$. But each thread is assigned a single Eulerian position $(x,y,z)$ in the neighborhood of $x(k)$, and are responsible for adding the appropriate value to $U(x,y,z)$ and $G(x,y,z)$ in this cell, as illustrated in Figure~2 b. In other words, CUDA blocks are built to have the same $5 \times 5 \times 5$ shape as the neighbourhood of a Lagrangian position. With this strategy, write operations into $U$ and $G$ variables are coalesced within one block.

In this case, different blocks concurrently write into $U(x,y,z)$ and $G(x,y,z)$ at the same cell. To avoid a resulting race conditions, the write operations are made atomic with help of the CUDA function \textit{atomicAdd()}. 

The pseudo-code of this kernel is written as follows:

\begin{algorithm}[H]  
\caption{The kernel ib\_force2: computation of $G(X, Y, Z)$ and $U(X, Y, Z)$ on all lattice cells $X, Y, Z$}  
\begin{algorithmic}  
\STATE ib\_force2(object, $U$, $G$, fixed point iteration)  
\STATE $k \gets blockIdx.x$\\  
   
\STATE $x(t),y(t),z(t) \gets $object's point $k$ transformed position\\  
   
\STATE $X \gets round(x(t)) + threadIx.x - 2$\\  
\STATE $Y \gets round(y(t)) + threadIx.y - 2$\\  
\STATE $Z \gets round(z(t)) + threadIx.z - 2$\\  
   
$weight \gets w(|x(t) - X| ) \times w(|y(t) - Y|) \times w(|z(t) - Z|)$\\  
\IF{Last fixed point iteration}  
    \STATE $G(X, Y, Z) \gets \Delta u(k) \times weight$  
\ELSE  
    \STATE $U(X, Y, Z) \gets U(X, Y, Z) + g(k)\times weight$\\  
\ENDIF  
\end{algorithmic}  
\end{algorithm} 
  
\begin{figure}[H] 
\centering  
\subfigure[\textit{ib\_force1} kernel]{ 
\centering   
\includegraphics[width=50 mm]{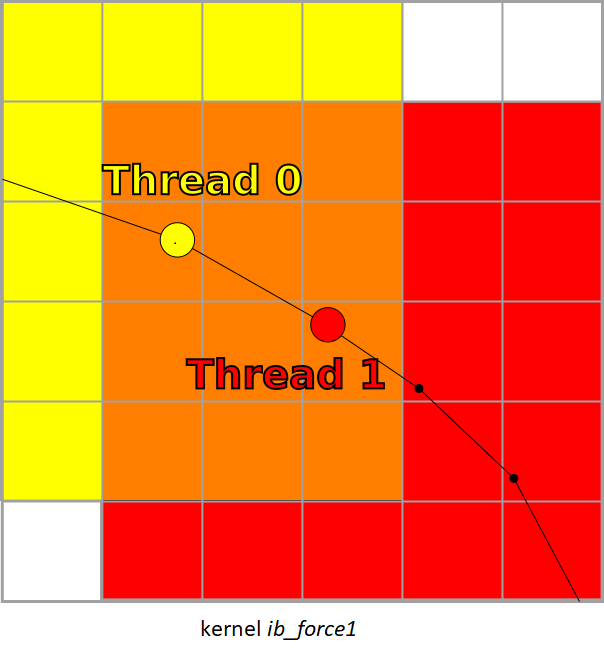}   
} 
\subfigure[\textit{ib\_force2} kernel]{ 
\centering   
\includegraphics[width=50 mm]{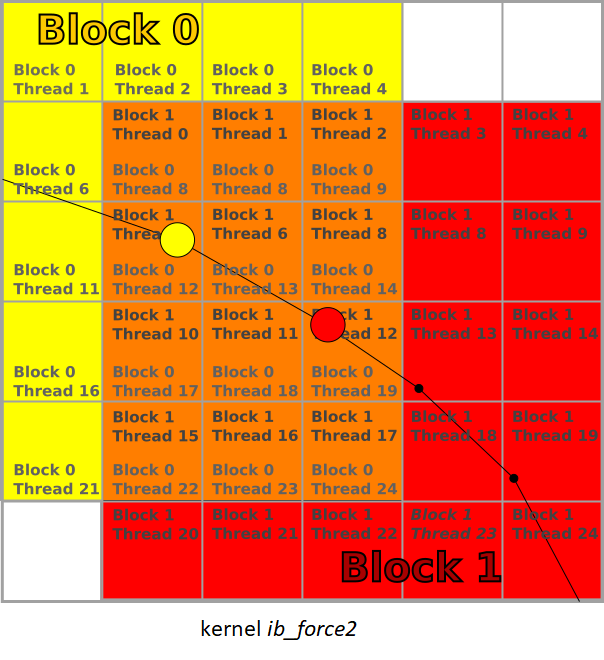}   
} 
\caption{{Schematic view of the attribution of parts of the computational domain to GPU. The grid represents the fluid lattice, while the black line stands for a piece of the object surface.}}   
\end{figure} 
  
\section{Optimisations}  
One of the problems that stem from the IBM is that the velocity $U$ needs to be precomputed prior to the collision phase, as it is needed for the computation of the IBM force. As a result, all cells populations need to be loaded twice, once to compute the velocity $U$ and once to compute the collision phase of the lattice Boltzmann. Our optimisation attempts are based on the observation that for the IBM, $U$ is required only in vicinity of the object surface and therefore can be precomputed in a domain of limited extent.

We tried two optimisation strategies:   
\begin{description}  
    \item[The box strategy], in which $U$ is precomputed only in a bounding box around all points on the object surface. Given that the object may move, the bounding box needs to be recomputed at every iteration. To achieve this efficiently in our algorithm, the assumption of rigid-body motion is used, and the bounding box of the object is computed only in the rest position. Then, at every iteration the linear transformation $M(t)$ is applied to the rest-position bounding box to obtain the current one.
  
    \item[The kernel strategy], in which $U$ is precomputed exactly for the points in which it will be needed, in the neighbourhood of the object surface. The implementation of this strategy is similar to the one of the kernel $ib\_force2$, using a CUDA block to compute $U$ in the neighbourhood of a Lagrangian point. The obtained domain is tighter than the one resulting from the box strategy. However, the same value of $U(x,y,z)$ is computed multiple times, since there are substantial overlaps between the neighborhoods of different points. As a result, this strategy is sometimes faster than the box strategy, and in some cases slower. 
\end{description}  
\begin{figure}[H] 
\centering  
\subfigure[The box strategy]{ 
\centering   
\includegraphics[width=50 mm]{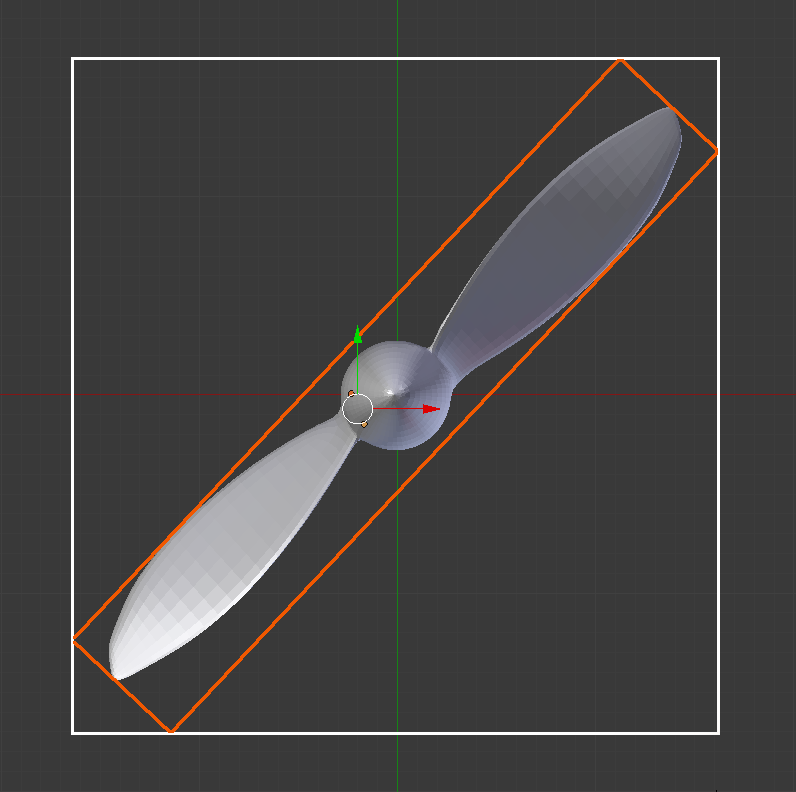}   
} 
\subfigure[The kernel strategy]{ 
\centering   
\includegraphics[width=70 mm]{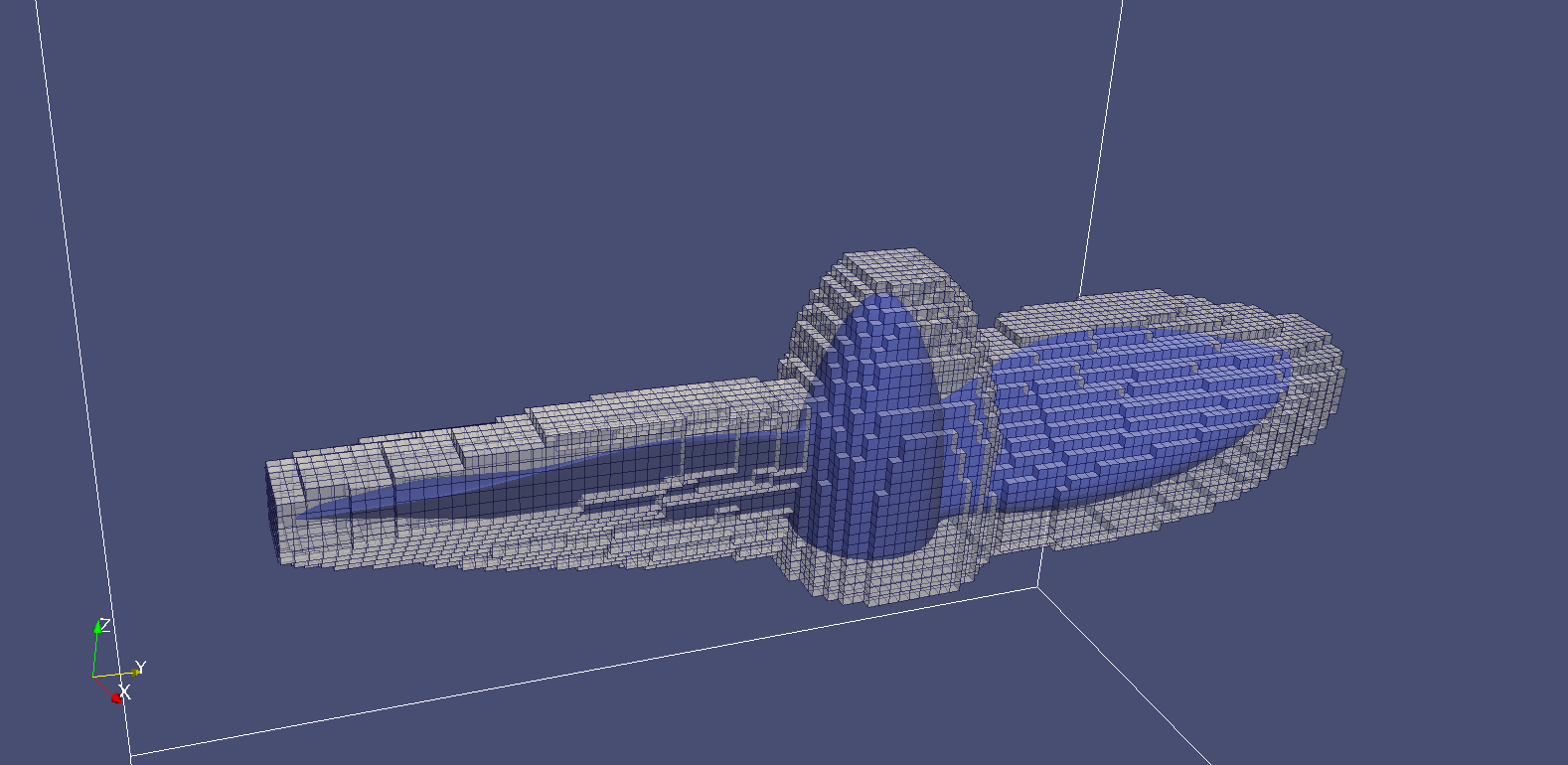}   
} 
\caption{Representation of the two implementation strategies for the \textit{lbm\_u\_start} kernel.}   
\end{figure} 
  
\section{Test cases}
\begin{figure}[H]   
\centering   
\caption{Visual representation of the test cases with 1, 6, and 18 propellers.}   
\centerline{%
\includegraphics[width=40 mm]{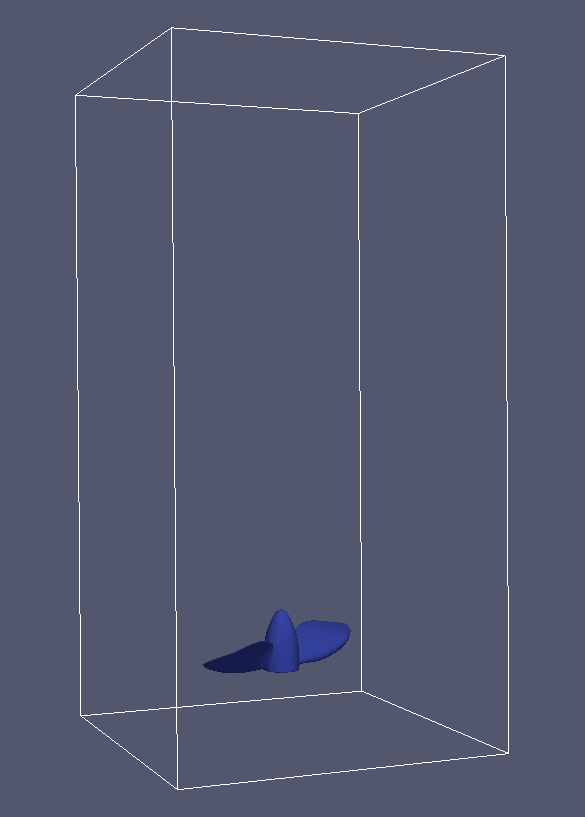}   
\includegraphics[width=40 mm]{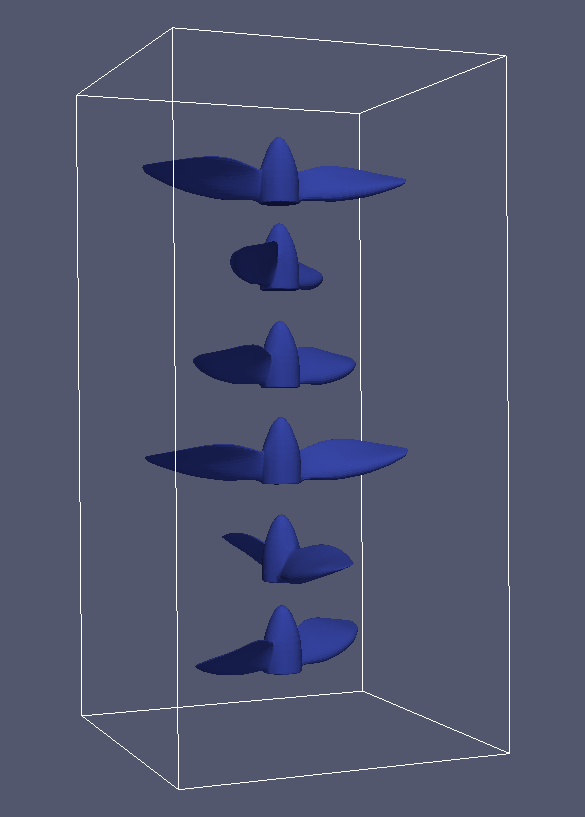}   
\includegraphics[width=40 mm]{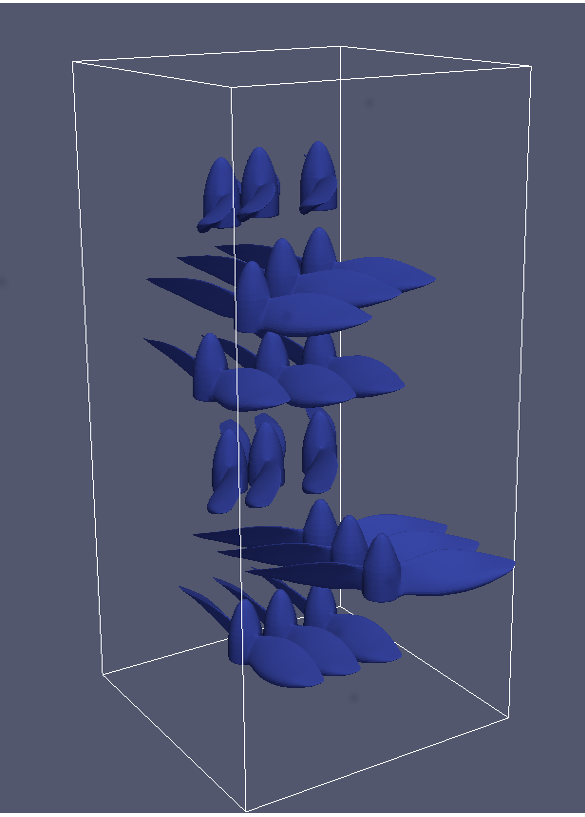}  
}%
\end{figure}  
  
Our test cases implement a rotating propeller, the geometry of which was created artistically, without any assumption on the use of the propeller in fluid engineering. The surface mesh of the propeller was built with help of the CAD functionality of the Blender software and consists of 3930 Lagrangian points. Before superimposing the surface mesh to the regular fluid mesh, it is rescaled to a size at which the area assigned to a Lagrangian point corresponds approximately to a 2D cross-section of a 3D fluid cell, in order to guarantee the accuracy of the IBM. The fluid volume is resolved by a lattice of $160 \times 160 \times 320$ cells, as shown in Figure~4. The benchmark cases are executed during 1000 iterations, and the performance of the code is asserted, as it is custom in the LBM community, by a time-averaged measure of million lattice-cell updates per second (Mlups). 
  
Both GPU and CPU tests were run on the parallel computer Baobab at the University of Geneva. The CPU is an Intel Xeon E5-2643 v3 CPU at 
3.40 GHz. The GPU is a NVIDIA Tesla P100-PCIE GPU with 3584 CUDA cores at 1.33 GHz. 
  
We ran 6 tests with one to six immersed and simultaneously rotating propellers. Each test was executed using both the kernel strategy and the box strategy, and the results are shown in Figure~5. In all six tests, the kernel strategy achieved better performances than the box strategy, and the difference in performance increased with the number of propellers. Our best measured performances correspond to 893 Mlups in the single-propeller and 650 Mlups for 6 propellers.

These performance values are similar to the ones obtained by other authors, and are more than an order of magnitude above the performances obtained on a CPU. The purpose of this article, however, is to show that our implementation strategy yields a substantial speedup compared to a CPU even with a much larger number of Lagrangian points, representing a situation of a dense arrangement of immersed objects. Indeed, one can consider such a situation to be in principle more favorable for a CPU rather than a GPU implementation, due to the frequent irregular memory traversal patterns.

For the sake of comparison, the CPU version of the code was executed with help of the high-performance LBM library Palabos, which used MPI parallelism to use all 12 cores available on the test CPU. Figure~6 shows a comparison of the two GPU versions and the Palabos CPU version for 1-6 propellers, and for an extreme case including 18 propellers. In the one-propeller case, the best performance is of 893 Mlups on GPU against 45.3 Mlups on the CPU (the GPU is 19.7 times faster), and in the 18-propeller case, the GPU yields 344.2 Mlups against 21.7 Mlups for the CPU (the GPU is 15.8 times faster). In the case of 18 propellers, the simulation uses 70740 Lagrangian points, and the neighborhoods of the immersed objects fill a substantial volume of the fluid. In this case, the kernel strategy loses its advantage against the box strategy, and both yield approximately the same performance.
  
\begin{figure}[H]   
\centering   
\includegraphics[width=100 mm]{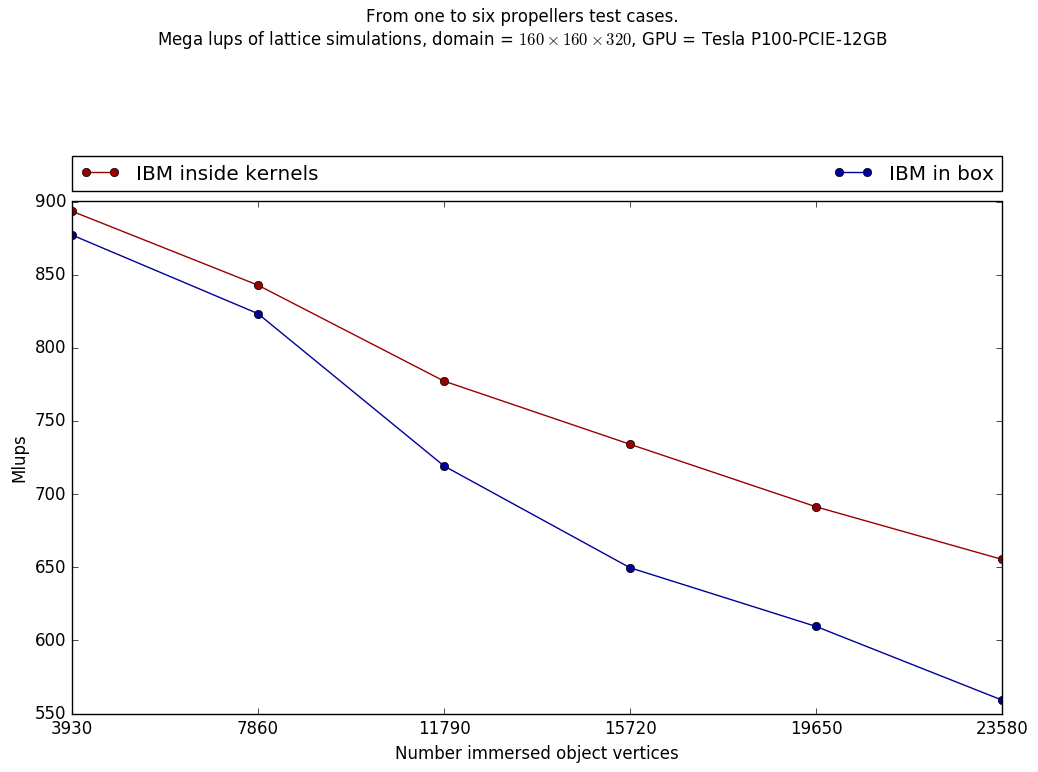}   
\caption{Performance measurements for 1 to 6 rotating propellers.}   
\end{figure}\begin{figure}[H]   
\centering   
\includegraphics[width=100 mm]{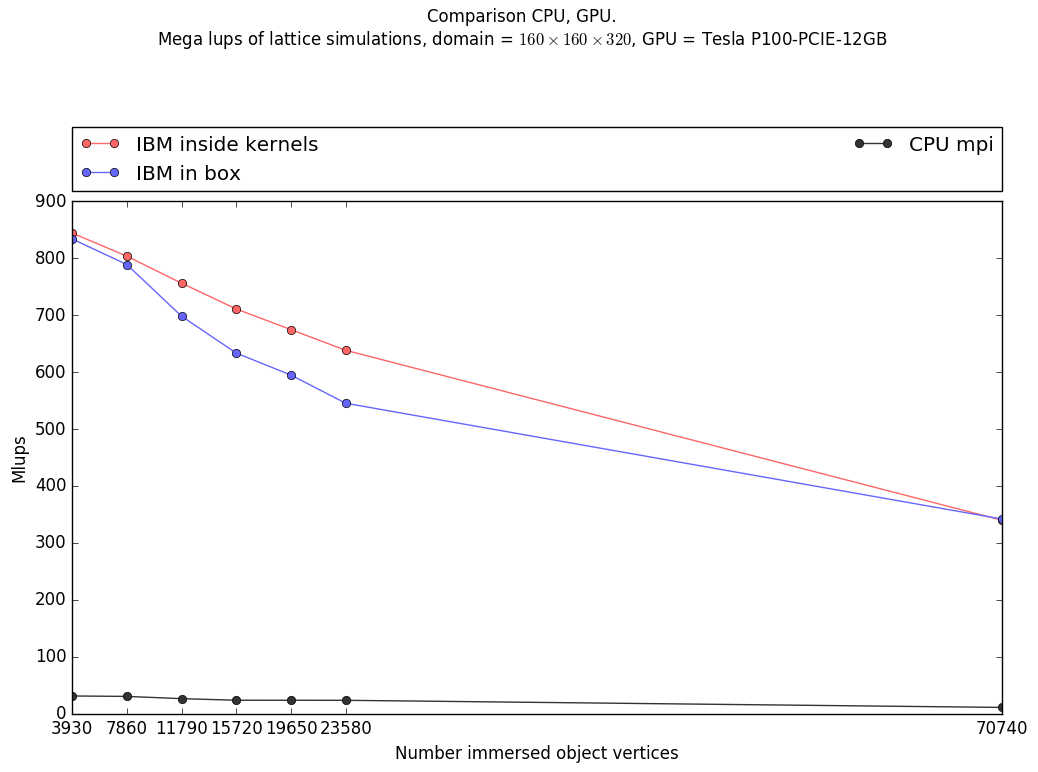}   
\caption{Performance comparison between GPU and CPU executions.}   
\end{figure}

\section{Conclusion}  
In this article, we present a GPU implementation of an immersed-boundary LBM, capable of simulating moving, immersed rigid objects substantially faster than a CPU implementation of the same problem. Similar to other publications in the field, the implementation yields a speedup of a factor 20, approximately, of the GPU against the CPU. But unlike other articles, we consider cases in which multiple moving immersed objects are densely packed inside the fluid, in which the advantage of the GPU against the CPU remains substantial, with a speed up of approximately 15.

The GPU algorithm is split into sections which, individually, are straightforward and relatively similar to their sequential counterpart. The test runs show that the overall performance of the implementation decreases as the number of Lagrangian points is increased. This is unsurprising, as the number of Lagrangian points is superior to the number of available GPU cores, and is also compatible with the observation that the IBM algorithm is subject to a limited memory bandwidth, as the memory accesses for this algorithm are very often uncoalesced. Nevertheless, the GPU retains a speedup of an order of magnitude compared to the CPU, and should therefore be considered an almost compulsory choice, even for non trivial problems including densely packed moving objects.

Although it would have been interesting to test the scaling of the implementation to an even larger number of Lagrangian points, this was not possible due to the limited amount of memory on the tested NVidia GPU. More generally, there is a need to generalize the proposed algorithm to a multi-GPU context to access larger domains, a project which we reserve for future work.
  

\end{document}